# Study of fluid displacement in three-dimensional porous media with an improved multi-component pseudopotential lattice Boltzmann method

M. Sedahmed*[1], R. C. V. Coelho[2,3], N. A. M. Araújo[2,3], E. M. Wahba[1], H. A. Warda[1]

*(1) Mechanical Engineering Department, Alexandria University, Alexandria 21544, Egypt.*
*(2) Centro de Física Teórica e Computacional, Faculdade de Ciências, Universidade de Lisboa, 1749-016 Lisboa, Portugal.*
*(3) Departamento de Física, Faculdade de Ciências, Universidade de Lisboa, 1749-016 Lisboa, Portugal.*
**Corresponding author email: mahmoud.sedahmed@alexu.edu.eg*

## ABSTRACT

We generalize to three dimensions (3D) a recently developed improved multi-component pseudopotential lattice Boltzmann method and analyze its applicability to simulate flows through realistic porous media. The model is validated and characterized via benchmarks, and we investigate its performance by simulating the displacement of immiscible fluids in 3D geometries. Two samples are considered, namely, a pack of spheres obtained numerically, and a Bentheimer sandstone rock sample obtained experimentally. We show that, with this model it is possible to simulate realistic viscosity ratios, to tune surface tension independently and, most importantly, to preserve the volume of trapped fluid. We also evaluate the computational performance of the model on the Graphical Processing Unit (GPU) and mention the implemented optimizations to increase the computational speed and reduce the memory requirements.

## 1. INTRODUCTION

Multiphase flow through porous medium occurs in several industrial and environmental processes, such as oil and gas production and carbon dioxide capture and storage (CCS). Such applications have a significant economic and environmental impact. Most of the world's energy is generated from oil and gas which are extracted from underground porous rocks (reservoirs). Also, one of most promising techniques to reduce global emissions from carbon dioxide is collect it from major sources and store it in underground depleted reservoirs. Hence, it is vital to understand the underlying physics involved in these processes [1].

For years, the fluid displacement processes in porous media were described at the macroscopic scale (meters to kilometers) using the well-known Darcy equation. That approach relied on averaged effective properties such as porosity and permeability. However, much underlying physics occur at the pore scale (micrometers to millimeters) where flow properties on this scale ultimately determine the larger scale properties. Advances in pore-scale physics provide ways to fill the gap between these scales. Developing numerical tools that could mimic pore-scale physics is essential to fill that gap.

For instance, understanding and optimizing oil and gas recovery from reservoirs often includes reservoir simulations on a larger scale to evaluate different recovery scenarios and select the most economic technique. Some key inputs for these simulations are extracted from the results of laboratory experiments that mimic fluids flow on the pore scale. Since these experiments are often lengthy and costly, they are carried out in a limited manner and take several months or a year to provide meaningful results. Hence, having a numerical model that could mimic those laboratory experiments could potentially reduce the cost and time dramatically [2]. Moreover, it could provide a better understanding of the significance of fluid properties on the pore scale flow.

Over the past few decades, the lattice Boltzmann method (LBM) has emerged as a powerful numerical tool for fluid flow simulation [3] [4] [5] [6] [7] [8] [9] [10] [11] [12]. Several LBM multiphase models were introduced in the literature, with the most widely used models being the color gradient model [13], the free energy model [14], and the pseudopotential model [15]. In this work, we employ the multi-component (MC) pseudopotential model due to its simplicity [10] [16]. The MC pseudopotential model is also widely used to study immiscible fluid displacement in porous media. One of the early works that employed the pseudopotential model for that purpose is the work by Chen and Martys [17]. They used the pseudopotential model to simulate drainage and imbibition processes in a three-dimensional (3D) porous medium reconstructed from microtomography images of Fontainebleau sandstones. Pan *et al.* [18] simulated fluids flow in a synthetic packing, and they compared numerical and experimental results. Porter *et al.* [19] investigated hysteresis in the relationship between capillary pressure $P_c$ and wetting phase saturation $S_w$ for drainage and imbibition processes. Warda *et al.* [20]

simulated the primary drainage and imbibition displacement processes in a 2D generated heterogeneous porous medium and showed that the MC model captures the capillary pressure bump phenomenon due to heterogeneity of the porous medium. Fager et al. [21] used the pseudopotential LBM model in a digital SCAL workflow to directly simulate an enhanced oil recovery (EOR) technique named water alternating gas (WAG) on the pore scale. Nemer *et al.* [22] investigated the relative importance of the wettability-altered fraction, the degree of wettability alteration, the accurate contact angle assignment, and their effects on relative permeability and fluid configurations using the MC model. Liu *et al.* [23] reviewed recent reports that demonstrate the power of LBM methods in terms of single and multiphase fluid flow in porous media and demonstrated the great potential in solving advanced waterflooding problems in complex porous media via coupling the LBM with other reactive transport solvers. Zheng *et al.* [24] studied the spontaneous imbibition phenomenon of fracturing fluid into shale matrix using an improved MC pseudopotential model and they were able to provide a deeper understanding of the phenomenon and its impacts. Fager *et al*. [25] explored the impact of different parameters such as capillary number, wetting condition, viscosity ratio, and driving mechanism on the oil/water displacement process in 3D models of porous rock.

The are several limitations in the original pseudopotential model. For instance, the viscosity ratio is limited (~10), the tuning of the surface tension depends on other quantities [8], and there is a non-negligible change in the volume of trapped fluids over time, which is non-physical [26]. Several enhancements were introduced to the model to address these limitations as, for instance, enhancements to the collision model [8] [9]. Also, some enhanced models relied on mixing various enhancements from the literature to form a new model [27] [28] [29].
Moreover, the original wetting boundary condition of the model resulted in non-physical effects near the fluid-solid interface. However, those could be suppressed using an improved wetting boundary condition, as shown in Ref. [30].

Here, we generalize a recently developed pseudopotential LBM model in three dimensions used for the simulation of immiscible fluids displacement [16]. The model is based on a modified explicit forcing (EF) scheme combined with the single relaxation time (SRT) approximation of the collision operator in the lattice Boltzmann equation (LBE) [31] [32]. That model achieved high viscosity ratios (~250) with the simple SRT approximation [31]. The model also employs a set of treatments for boundary conditions that were shown to aid in suppressing the non-physical behavior of the pseudopotential model, namely, a reduction in the volume of trapped fluids in porous media [16]. We use the generalized model in simulating immiscible fluid displacement in 3D realistic geometries of actual samples of porous media. We also investigate the performance in terms of capturing the physical behaviors of the simulated system and in terms of computational performance on the graphical processing unit (GPU) as well.

This paper is organized as follows. In section 2, we present the three-dimensional version of the improved multi-component pseudopotential model and boundary conditions. Section 3 provides numerical tests and calibration for the model using some widely used benchmarks from the literature. In section 4, we present the simulation results of immiscible fluids displacement in different examples of porous media, namely, a pack of spheres and a sample of Bentheimer sandstone rock. In section 5, we discuss the performance of the code used to implement the model on GPU using several optimization techniques from the literature. In section 6, we draw some conclusions and discuss possible follow ups.

## 2. METHODOLOGY

### *Multi-component pseudopotential lattice Boltzmann method*

We employ a three-dimensional version of a model developed in our previous work [16], which is summarized below. The model is based on the MC pseudopotential lattice Boltzmann method with the modified explicit forcing (EF) scheme and the SRT approximation of the collision operator [31]. The discretized lattice Boltzmann equation (LBE) for this model is written as

$$f_i^{(\sigma)}(\boldsymbol{x}+\boldsymbol{e}_i\,\Delta t, t+\Delta t) - f_i^{(\sigma)}(\boldsymbol{x},t) = -\frac{1}{\tau_s}\left(f_i^{(\sigma)}(\boldsymbol{x},t) - f_i^{(\sigma),eq}(\boldsymbol{x},t)\right) + \Delta t\left(1-\frac{1}{2\,\tau_s}\right) S_i^{(\sigma),F}(\boldsymbol{x},t), \quad (1)$$

where, $f_i^{(\sigma)}$ is the particle distribution function at position $(\boldsymbol{x})$ and time $(t)$, $\sigma$ is the fluid component number $(\sigma = 1, \ldots n)$. We consider only two fluid components, i.e., $n = 2$. $i$ is the lattice direction that belongs to the selected lattice arrangement; Here, we use the $D_3Q_{19}$ lattice arrangement for 3D simulations $(i = 0, 1, \ldots 18)$. $f_i^{(\sigma),eq}$ is the equilibrium distribution function, and it reads

$$f_i^{(\sigma),eq}(\boldsymbol{x}) = w_i \rho^{(\sigma)}(\boldsymbol{x})\left[1 + \frac{\boldsymbol{e}_i\cdot\boldsymbol{u}^{eq}(\boldsymbol{x})}{c_s^2} + \frac{\left(\boldsymbol{e}_i\cdot\boldsymbol{u}^{eq}(\boldsymbol{x})\right)^2}{2\,c_s^4} - \frac{\boldsymbol{u}^{eq}(\boldsymbol{x})^2}{2\,c_s^2}\right], \quad (2)$$

where, $c_s$ is the lattice sound speed, and it is defined as $c_s = \frac{\Delta x}{\sqrt{3}\,\Delta t}$. $\Delta x$ and $\Delta t$ are commonly chosen as unity in the LBM. $w_i$ and $\boldsymbol{e}_i$ are the lattice weights and the discrete velocity vector in the $i^{th}$ direction, respectively, which are given by

$$\boldsymbol{e}_i = \begin{bmatrix} 0 & 1 & -1 & 0 & 0 & 0 & 0 & 1 & -1 & 1 & -1 & 1 & -1 & 1 & -1 & 0 & 0 & 0 & 0 \\ 0 & 0 & 0 & 1 & -1 & 0 & 0 & 1 & 1 & -1 & -1 & 0 & 0 & 0 & 0 & 1 & -1 & 1 & -1 \\ 0 & 0 & 0 & 0 & 0 & 1 & -1 & 0 & 0 & 0 & 0 & 1 & 1 & -1 & -1 & 1 & 1 & -1 & -1 \end{bmatrix},$$

$$w_i = \begin{bmatrix} \frac{1}{3} & \frac{1}{18} & \frac{1}{18} & \frac{1}{18} & \frac{1}{18} & \frac{1}{18} & \frac{1}{18} & \frac{1}{36} & \frac{1}{36} & \frac{1}{36} & \frac{1}{36} & \frac{1}{36} & \frac{1}{36} & \frac{1}{36} & \frac{1}{36} & \frac{1}{36} & \frac{1}{36} & \frac{1}{36} & \frac{1}{36} \end{bmatrix}.$$

The macroscopic fluid component density, velocity, and mixture (total) density are given by

$$\rho^{(\sigma)}(\boldsymbol{x}) = \sum_i f_i^{(\sigma)}(\boldsymbol{x}), \quad (3)$$

$$\boldsymbol{u}^{(\sigma)}(\boldsymbol{x}) = \frac{1}{\rho^{(\sigma)}(\boldsymbol{x})}\sum_i \boldsymbol{e}_i f_i^{(\sigma)}(\boldsymbol{x}), \quad (4)$$

$$\rho(\boldsymbol{x}) = \sum_\sigma \rho^{(\sigma)}(\boldsymbol{x}). \quad (5)$$

The system relaxation time $\tau_s$ is defined as

$$\tau_s(\boldsymbol{x}) = \frac{\sum_\sigma \rho^{(\sigma)}(\boldsymbol{x})\, \nu^{(\sigma)}}{\rho(\boldsymbol{x})\, c_s^2\, \Delta t} + \frac{1}{2}, \tag{6}$$

where, $\nu^{(\sigma)}$ is the kinematic viscosity of the $\sigma^{th}$ fluid component, and it is determined by

$$\nu^{(\sigma)} = c_S^2 \left(\tau^{(\sigma)} + \frac{1}{2}\right) \Delta t, \tag{7}$$

where, $\tau^{(\sigma)}$ is the relaxation time of the $\sigma^{th}$ fluid component. The system relaxation time $\tau_s$ is smooth at the interface between the two fluids in order to make the collision model more stable.

The equilibrium velocity $\boldsymbol{u}^{eq}$ reads

$$\boldsymbol{u}^{eq}(\boldsymbol{x}) = \frac{\sum_\sigma \frac{\rho^{(\sigma)}(\boldsymbol{x})\, \boldsymbol{u}^{(\sigma)}(\boldsymbol{x})}{\tau_s(\boldsymbol{x})}}{\sum_\sigma \frac{\rho^{(\sigma)}(\boldsymbol{x})}{\tau_s(\boldsymbol{x})}}. \tag{8}$$

Also, the common physical velocity of the two fluid components is defined as $\boldsymbol{u}(\boldsymbol{x}) = \boldsymbol{u}^{eq}(\boldsymbol{x})$.

The forcing term in the LBE equation $S_i^{(\sigma),F}$ is defined as follows:

$$S_i^{(\sigma),F}(\boldsymbol{x}) = w_i \left(\frac{\boldsymbol{e}_i - \boldsymbol{u}^{eq}(\boldsymbol{x})}{c_S^2} + \frac{\boldsymbol{e}_i \cdot \boldsymbol{u}^{eq}(\boldsymbol{x})}{c_S^4} \boldsymbol{e}_i\right) \cdot \boldsymbol{F}^{(\sigma),tot}(\boldsymbol{x}), \tag{9}$$

where, $\boldsymbol{F}^{(\sigma),tot}$ is the total force exerted on the $\sigma^{th}$ fluid component and it is defined for the pseudopotential model by,

$$\boldsymbol{F}^{(\sigma),tot} = \boldsymbol{F}^{(\sigma),f-f} + \boldsymbol{F}^{(\sigma),f-s} + \boldsymbol{F}^{(\sigma),b}, \tag{10}$$

where, $\boldsymbol{F}^{(\sigma),f-f}$ is the fluid-fluid interaction force of the $\sigma^{th}$ fluid component, and it is determined by,

$$\boldsymbol{F}^{(\sigma),f-f}(\boldsymbol{x}) = -G_{coh}^{(\sigma\bar{\sigma})} \psi^{(\sigma)}(\boldsymbol{x}) \sum_i w_i\, \psi^{(\bar{\sigma})}(\boldsymbol{x} + \boldsymbol{e}_i)\boldsymbol{e}_i, \tag{11}$$

where, $\psi^{(\sigma)}$ is the pseudopotential of the $\sigma^{th}$ fluid component. Here we define it as $\psi^{(\sigma)} = \rho^{(\sigma)}$, $G_{coh}^{(\sigma\bar{\sigma})}$ is the fluid-fluid cohesion strength, and it is used to tune the fluid-fluid interaction force. This parameter controls the segregation between the two components, and it is used for tuning the surface tension in the pseudopotential model. It should be noted that only the inter-component interaction force is considered in this work $(\sigma \neq \bar{\sigma})$ while the intra-component interaction force is neglected $\left(G_{coh}^{(\sigma\sigma)} = 0\right)$. We set $G_{coh}^{(\sigma\bar{\sigma})} = 3.5$ in our simulations.

We point out that at boundary points where non-periodic boundary conditions are applied (e.g., Dirichlet boundary condition), there will be an imbalance in the force summation term of Eq. (11) as the neighbor points located at $(\boldsymbol{x} + \boldsymbol{e}_i)$ will be located outside the domain boundaries. Hence, the treatment presented in Ref. [16] is used to provide this balance. For example, at the left boundary, summation terms of pseudopotential in five directions would be missing, i.e., $i = 2,8,10,12,14$. The pseudopotential values in the neighbor lattice points adjacent to these directions would be determined using as follows:

$$\psi^{(\bar{\sigma})}(\boldsymbol{x}+\boldsymbol{e}_2) = \psi^{(\bar{\sigma})}(\boldsymbol{x}+\boldsymbol{e}_1),$$
$$\psi^{(\bar{\sigma})}(\boldsymbol{x}+\boldsymbol{e}_8) = \psi^{(\bar{\sigma})}(\boldsymbol{x}+\boldsymbol{e}_7),$$
$$\psi^{(\bar{\sigma})}(\boldsymbol{x}+\boldsymbol{e}_{10}) = \psi^{(\bar{\sigma})}(\boldsymbol{x}+\boldsymbol{e}_9),$$
$$\psi^{(\bar{\sigma})}(\boldsymbol{x}+\boldsymbol{e}_{12}) = \psi^{(\bar{\sigma})}(\boldsymbol{x}+\boldsymbol{e}_{11}),$$
$$\psi^{(\bar{\sigma})}(\boldsymbol{x}+\boldsymbol{e}_{14}) = \psi^{(\bar{\sigma})}(\boldsymbol{x}+\boldsymbol{e}_{13}).$$

The external body forces exerted on the $\sigma^{th}$ fluid component (e.g., gravity) could be added using the term $\boldsymbol{F}^{(\sigma),b}$**.** Several methods in the literature were introduced to incorporate these forces. One is referred to Ref. [8] for a review of different techniques.

In the present pseudopotential model, the pressure is determined as follows:

$$P = c_s^2 \sum_{\sigma} \rho^{(\sigma)} + \frac{c_s^2\,\Delta t}{2} G_{coh}^{(\sigma\bar{\sigma})} \sum_{\sigma,\bar{\sigma}} \rho^{(\sigma)} \rho^{(\bar{\sigma})}. \tag{12}$$

We express our results in lattice units (l.u.) which is a set of units defined such that the lattice spacing ($\Delta x$), the time step ($\Delta t$), and density ($\rho$) are equal unity ($\Delta x = \Delta t = \rho = 1$). Simulation results in lattice units could be converted to physical units using relevant dimensionless numbers (e.g., Reynolds number, capillary number, viscosity ratio). This conversion of units is based on the law of similarity where two fluid flows that have the same dimensionless numbers provide the same physics upon a simple scaling [6]. Thus, we only need to ensure that the relevant dimensionless numbers are the same in the two systems of units.

We use the improved virtual solid density (IVSD) scheme for the wetting boundary condition [30] [16]. The fluid-solid interactions are modeled in a similar manner as in Eq. (11), and it is re-written as

$$\boldsymbol{F}^{(\sigma),f-s}(\boldsymbol{x}) = -G_{coh}^{(\sigma\bar{\sigma})} \psi^{(\sigma)}(\boldsymbol{x}) \sum_i w_i\, s^{(\bar{\sigma})}(\boldsymbol{x}+\boldsymbol{e}_i)\boldsymbol{e}_i, \tag{13}$$

$s^{(\bar{\sigma})}$ is the solid pseudopotential of the $\bar{\sigma}$ fluid component, and it is defined as follows,

$$s^{(\sigma)}(\boldsymbol{x}) = \phi(\boldsymbol{x})\, \tilde{\rho}^{(\sigma)}(\boldsymbol{x}) \tag{14}$$

where, $\phi$ is a binary switch function that equals 0 for a fluid node and 1 for a solid node, $\tilde{\rho}^{(\sigma)}$ is the virtual solid density, and it is defined using the averaged density of the fluid as

$$\tilde{\rho}^{(\sigma)}(\boldsymbol{x}) = \chi^{(\sigma)} \frac{\sum_i w_i \rho^{(\sigma)}(\boldsymbol{x}+\boldsymbol{e}_i)\big(1-\phi(\boldsymbol{x}+\boldsymbol{e}_i)\big)}{\sum_i w_i \big(1-\phi(\boldsymbol{x}+\boldsymbol{e}_i)\big)}. \tag{15}$$

$\chi^{(\sigma)}$ is a factor that controls the wettability of the $\sigma$ fluid component. If the parameter $\chi^{(\sigma)}$ is unity, the fluid component has neutral wetting ($\theta = 90°$). To control the wettability of the fluid components, the parameter $\chi^{(\sigma)}$ is selected as follows (only two components are considered in this work, i.e., $\sigma = 1, 2$):

$$\chi^{(1)} = 1 + \xi,$$
$$\chi^{(2)} = 1 - \xi,$$

where, fluid component-1 is forming the droplets, and fluid component-2 is the surrounding fluid. Fluid component-1 becomes non-wetting if $\xi < 0$ and wetting if $\xi > 0$. We use the grouping of fluid wettability based on the contact angle shown in Refs. [33] [16].

For inlet and outlet boundary conditions, we use the set of boundary conditions developed in Ref. [16], which is summarized for the 3D implementation as follows,

1- Apply Zou-He (ZH) method [34] or the non-equilibrium extrapolation method (NEEM) [6] for only one fluid component at both inlet and outlet boundaries. Namely, apply fixed density for fluid component-1 at the inlet and apply fixed density for fluid component-2 at the outlet.
2- Apply the following equation as an open boundary condition for the missing distribution function for the other fluid component at the boundary,

$$f^{(1)}_{2,8,10,12,14}|_{out,t_1} \xrightarrow{collision} f^{(1)}_{2,8,10,12,14}|_{out,t_2} \xrightarrow{streaming} f^{(1)}_{2,8,10,12,14}|_{out,t_1},$$
$$f^{(2)}_{1,7,9,11,13}|_{in,t_1} \xrightarrow{collision} f^{(2)}_{1,7,9,11,13}|_{in,t_2} \xrightarrow{streaming} f^{(2)}_{1,7,9,11,13}|_{in,t_1},$$

where, $t_1$ is the current time step and $t_2$ is the new time step. It should be noted that since the missing distribution functions are keeping their values during the streaming step, they will maintain the initial condition values till the end of the simulation.

## 3. NUMERICAL TESTS

To validate and characterize the model in three dimensions, we performed the same benchmark tests which were considered in Ref. [16] for the 2D version of the model. Some of these benchmarks were also considered to calibrate the model parameters for other simulations, such as surface tension and contact angle.

### *3.1. Laplace test*

This test is carried out to verify that the model obeys the Laplace law in three dimensions $\left(\Delta P = \frac{2\gamma}{R}\right)$, where $\Delta P$ is the difference between the pressure inside and outside the droplet, $\gamma$ is the surface tension, and $R$ is the droplet radius.

We carried out 3D simulations with a droplet of fluid component-1 placed in a fully periodic domain of fluid component-2. The domain size was 128 x 128 x 128, and the initial droplet size was 30. The density field was initialized using the following equation,

$$\rho^{(\sigma)}(\boldsymbol{x}) = \frac{\rho^{(\sigma)}_{in} + \rho^{(\sigma)}_{out}}{2} - \frac{\rho^{(\sigma)}_{in} - \rho^{(\sigma)}_{out}}{2} \times \tanh\left[\frac{2\left(\sqrt{(x-x_c)^2 + (y-y_c)^2 + (z-z_c)^2} - r_0\right)}{W}\right], \quad (16)$$

where, $\rho_{in}^{(\sigma)} = 1$, $\rho_{out}^{(\sigma)} = 0.027$, $r_0 = 30$ and $W = 5$. $x_c$, $y_c$, and $z_c$ are locations of the center of the domain. This initialization method reduces the numerical instability in the first few iterations of the simulation as it represents an initial density field with a diffusive interface between the fluid components. Three different viscosity ratios were simulated $\left(M = 1, M = 25, M = \frac{1}{25}\right)$ by setting the relaxation times as follows:

$$\tau^{(\sigma)} = \begin{cases} \tau^{(1)} = \tau^{(2)} = 1\,, & M = 1 \\ \tau^{(1)} = 3.5, \tau^{(2)} = 0.62\,, & M = 25 \\ \tau^{(1)} = 0.62, \tau^{(2)} = 3.5\,, & M = \dfrac{1}{25}. \end{cases}$$

In this work, we assume both fluid components have the same density; hence the viscosity ratio is defined as $M = \nu_1/\nu_2$

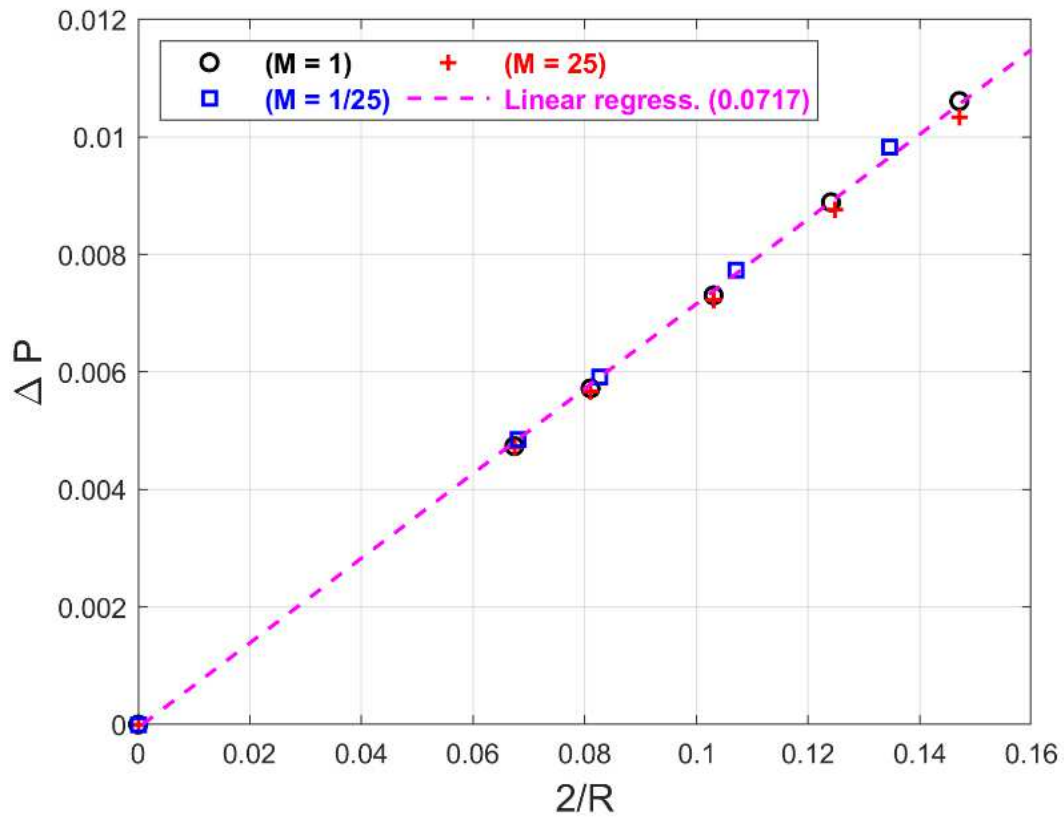

*Figure 1 Laplace test results. In the vertical axis is the pressure difference, while in the horizontal axis is twice the inverse of the droplet radius. Different symbols represent different viscosity ratio (black circle: M = 1, red cross: M = 25, blue square: M = 1/25). The dotted line in magenta represents the linear regression of data which gives a surface tension (slope) value* $\gamma = 0.071$ *corresponding to* $\left(G_{coh}^{(\sigma\bar{\sigma})} = 3.5\right)$.

After the simulations reached a steady state, the pressure inside and outside the droplet was measured and plotted against twice the inverse of the measured droplet radius, as shown in *Figure 1*. The relationship between $\Delta P$ and $\frac{2}{R}$ was indeed linear and thus satisfied the Laplace law. The surface tension was determined by the slope of the line. The original pseudopotential model suffered from the inability to set the surface tension independently from the viscosity ratio (e.g., Ref. [35]). However, the present model does not suffer from this limitation because, as shown in *Figure 1*, the surface tension does not depend on the selected viscosity ratio. Hence the model exhibits independent tuning of surface tension from the viscosity ratio.

### *3.2. Contact angle calibration*

This benchmark is conducted to calibrate the wetting boundary condition using the (IVSD) scheme introduced earlier (section 2) and to determine the relationship between the parameter $\xi$ and contact angle $\theta$ for the currently implemented model. In the pseudopotential model, the contact angle cannot be set directly into the model like, for instance, in the free energy model [6]. Hence, this characterization is essential to calibrate the model at a specific surface tension $\left(G_{coh}^{(\sigma\bar{\sigma})}\right)$ and properly set the desired contact angle in other simulations. We carried out 3D simulations where a semi-circular droplet of fluid component-1 with an initial droplet radius of 30 was initially placed on the surface of a solid wall. The domain size was 200 x 100 x 200 and periodic conditions in the x and z directions, while solid walls were located at the domain's top and bottom sides. The initial density values and $G_{coh}^{(\sigma\bar{\sigma})}$ were set as in section 3.1. The viscosity ratio was set as $M = 25$.

We specify the criteria for measuring the contact angle $(\theta)$ by the convention that the contact angle is measured through the denser fluid phase [1]. Since this work focused on an oil-water system, water is considered the denser fluid, and contact angle will be measured through the water, as in Refs. [16] [1]. The initial density field was initialized assuming that fluid component-1 (forming the droplet) is oil while fluid component-2 (surrounding fluid) is water. The contact angle of the droplet was measured in the steady state using the technique and geometrical relationships shown in Ref. [36], where the base width was measured five lattice units above the solid surface to avoid the influence of the surface on the measured interface. Additionally, a linear interpolation was used to determine the approximate location of the interface. Also, the halfway location of solid walls was considered [6].

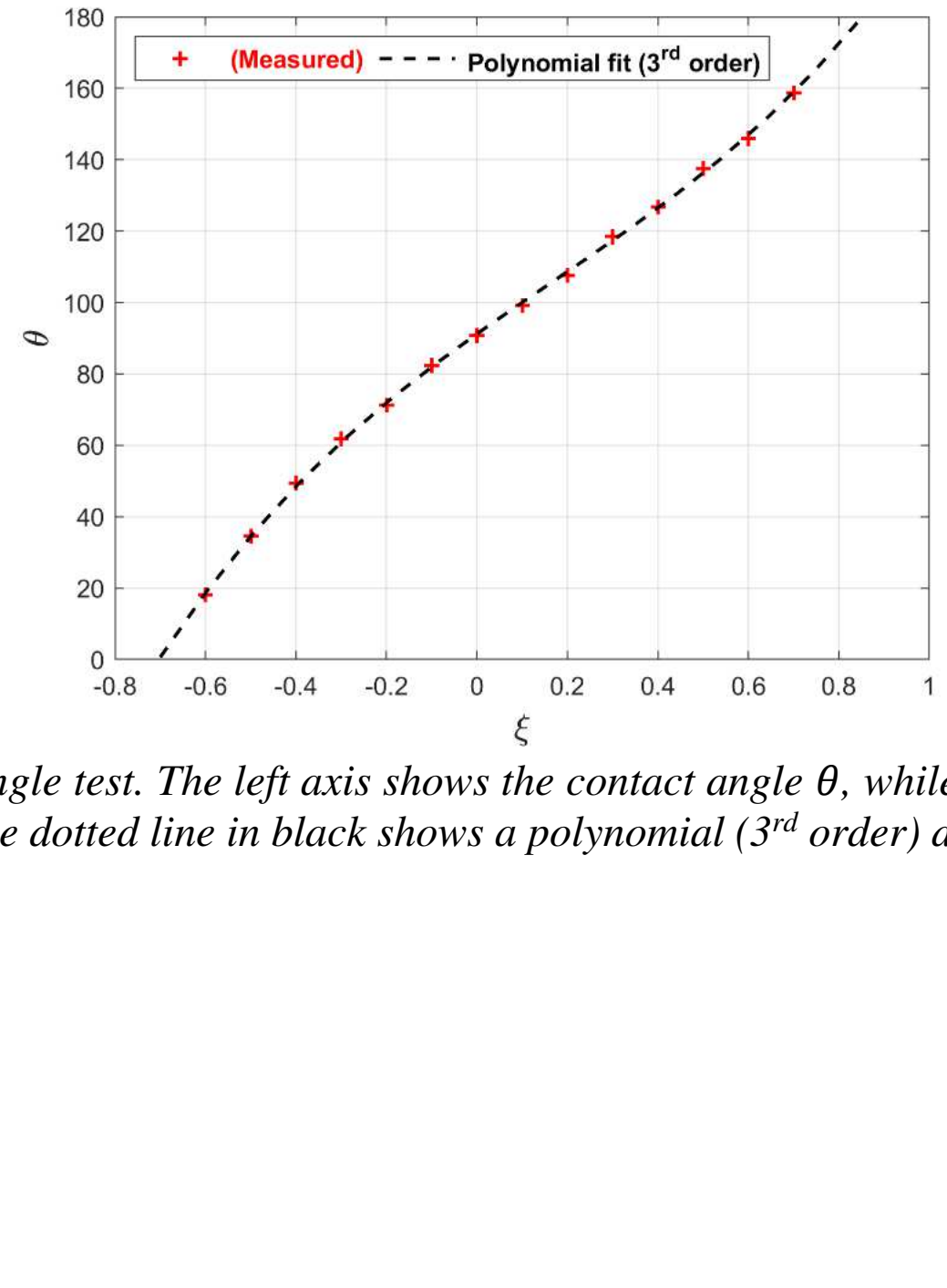

*Figure 2 Contact angle test. The left axis shows the contact angle $\theta$, while the bottom axis shows the parameter $\xi$. The dotted line in black shows a polynomial ($3^{rd}$ order) data fit.*

*Figure 2* shows the relationship between $\xi$ and contact angle $\theta$. We noted that the model produced the expected behavior in setting the wettability using the parameter $\xi$, where oil was non-wetting ($\theta < 90°$) for $\xi < 1$, and vice versa. A wide range of contact angles was successfully simulated, covering most practical applications. Moreover, an empirical equation could be deduced from the curves to represent the $\xi$-$\theta$ relationship: a third-order polynomial provided the best fit for our simulation results ($\theta = 47.4136\,\xi^3 - 23.4513\,\xi^2 + 89.7938\,\xi + 91.2324$). Snapshots of the simulation results at different contact angles are shown in *Figure 3*.

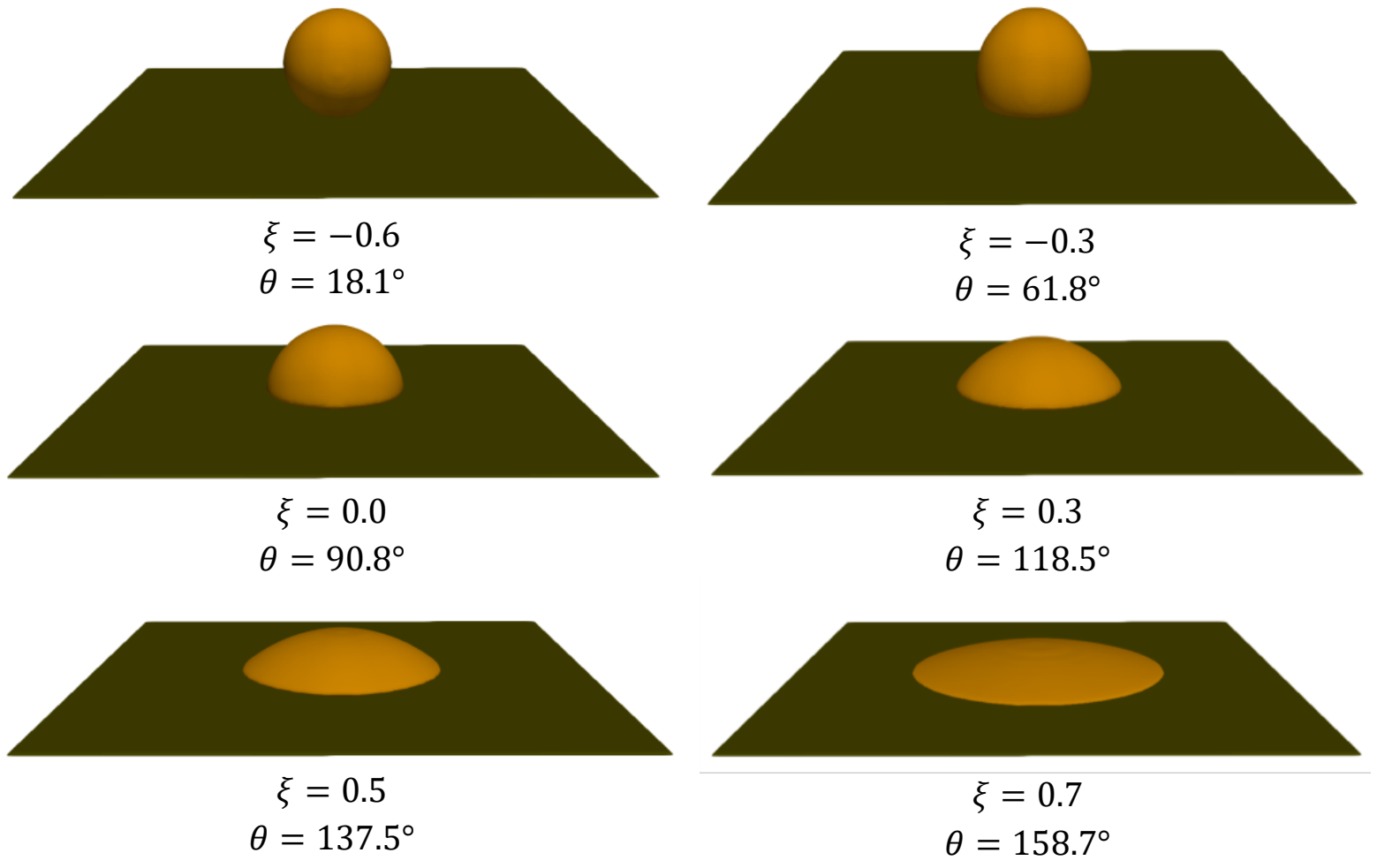

*Figure 3 Snapshots of contact angle test at different values of $\xi$ with corresponding values of contact angle $\theta$. Only oil is shown in the figures while water was hidden.*

## 4. POROUS MEDIA

Now we investigate the model performance and capabilities in the simulation of immiscible fluids displacement in three-dimensional realistic porous media. We show two examples that present different geometry types: a pack of spheres and a sample of a Bentheimer sandstone rock.

### *4.1. Immiscible fluids displacement in a pack of spheres*

We first consider a simple porous medium geometry consisting of a pack of equal size, hard spheres. The dataset representing the geometry was acquired from the Digital Rocks Portal under a project named "*Finney Packing of Spheres*" [37]. The original size of the dataset was 500 x 500 x 500. A subset with dimensions 200 x 200 x 200 was extracted and used in the present simulations. The porosity of the domain was 36%. A 2D slice of the used geometry is shown in *Figure 4*. The blue regions represent the pore network where the LBE equation is solved to describe the fluid flow, grey regions represent solid spheres, and red surfaces represent the fluid-solid (F-S) interface boundary between the pore and grain regions. The F-S surfaces are where the wetting boundary conditions are applied to characterize the wettability and are also used to apply the bounce-back boundary conditions.

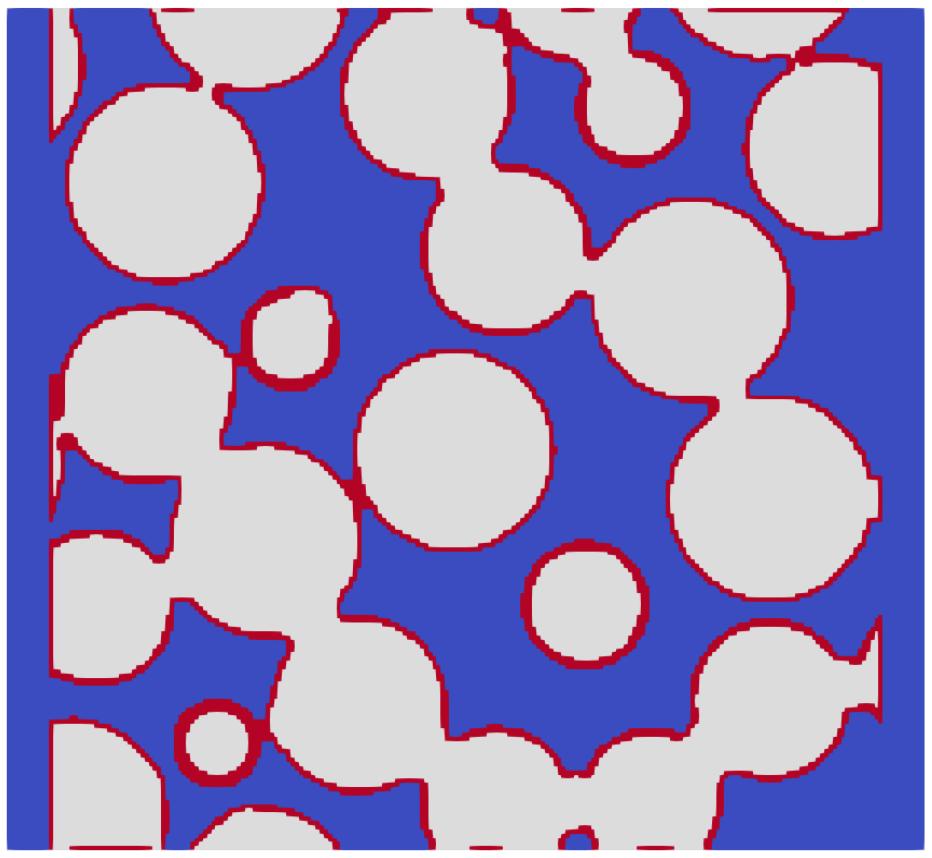

*Figure 4 Two-dimensional slice of the Finney Packing of Spheres located at z = 99. Different regions in the domain are identified as follows, blue: pore (fluid), grey: solid, red: fluid-solid interface boundary.*

Two sets of fluid layers were added at both the inlet (left) and outlet (right) boundary sides that represent pure reservoir fluid regions. These layers are used to inject and collect the fluids. The first (last) slice of the inlet (outlet) layer is used to implement the set of boundary conditions mentioned in section 2. Periodic boundary conditions were used in both the x and z directions. Initially, the domain is fully saturated with a wetting fluid (water), and only a few layers at the inlet side are saturated with a non-wetting fluid (oil). Hence, we start with the wetting phase saturation of $S_w = 1$. Water saturation $(S_w)$ was determined as the number of lattice points occupied by water divided by the total number of fluid lattice points. The viscosity ratio was set as $M = 25$ to represent a realistic viscosity ratio between oil and water. The initial density values were set as in section 3.1. The contact angle was set as $\theta \cong 70°$ ( $\xi = -0.2$).

We simulate a primary drainage displacement process [1] where the non-wetting fluid (oil) is displacing a wetting fluid (water) and gradually saturating the domain. The capillary pressure $P_c$ is considered the difference between the inlet and outlet pressures $P_c = P_{in} - P_{out}$. Inlet and outlet pressures are specified using Dirichlet boundary conditions via the non-equilibrium extrapolation (NEEM) method, where the inlet density was gradually increased while the outlet density was kept constant.

The primary drainage simulation results are shown in *Figure 5*. Several essential features could be observed. The non-wetting fluid (oil) did not start invading the domain until the capillary pressure exceeded a certain threshold that is known as the critical capillary pressure $(P_{c,t})$ which in this case was $P_{c,t} \cong 0.005$ at critical oil saturation $S_{o,c} \cong 0.05$. Then, as the capillary pressure was increased, the non-wetting fluid (oil) kept invading the domain gradually and the $P_c$-$S_w$ relation curve followed a typical curve shape for this case [38]. The wetting fluid (water) was displaced out of the domain until a saturation value was achieved despite the increase in the capillary pressure, which in this case was $S_{w,i} \cong 0.2$. The achieved water saturation $S_{w,i}$ indicated that the present model preserved the entrapped fluid volume. The current model was shown to have a similar advantage in two-dimensional cases as in Ref. [16], and it is shown in this study that the model keeps the same benefit in three-dimensional cases as well. Snapshots for the primary drainage displacement process are shown in *Figure 6*.

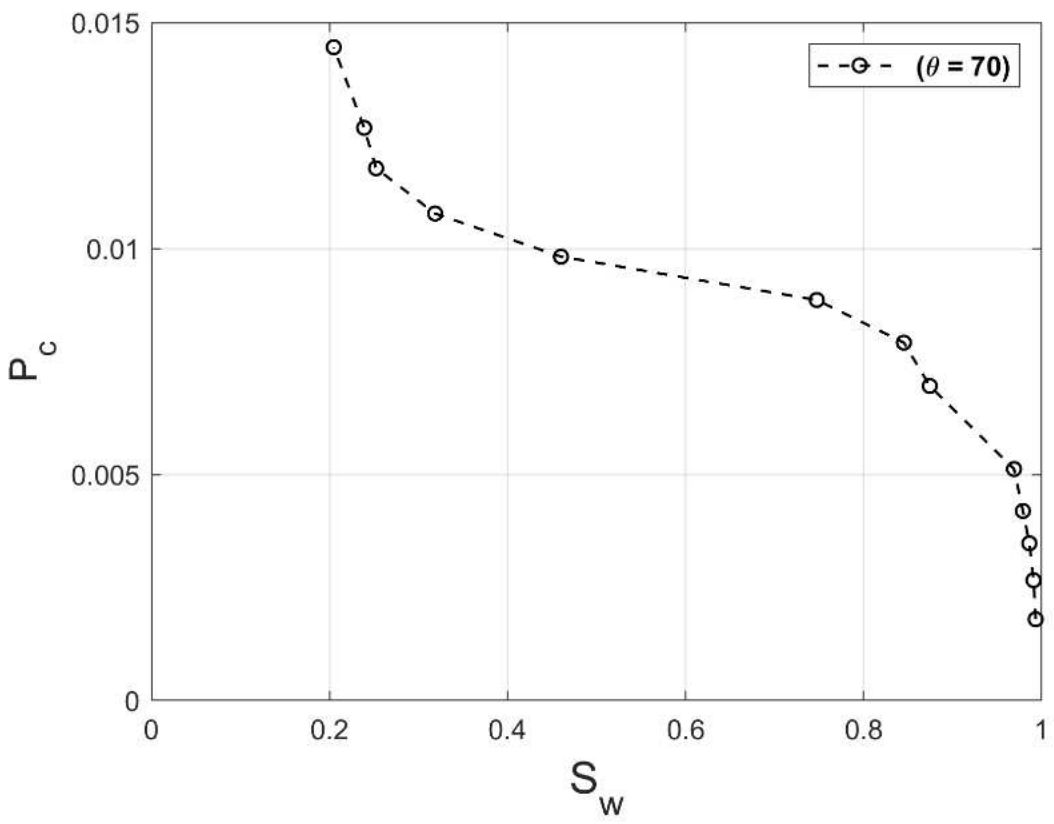

*Figure 5 . Capillary pressure* $(P_c)$ *– water saturation* $(S_w)$ *relationship for the primary drainage displacement process in the Finney pack of spheres at contact angle* $\theta = 70°$.

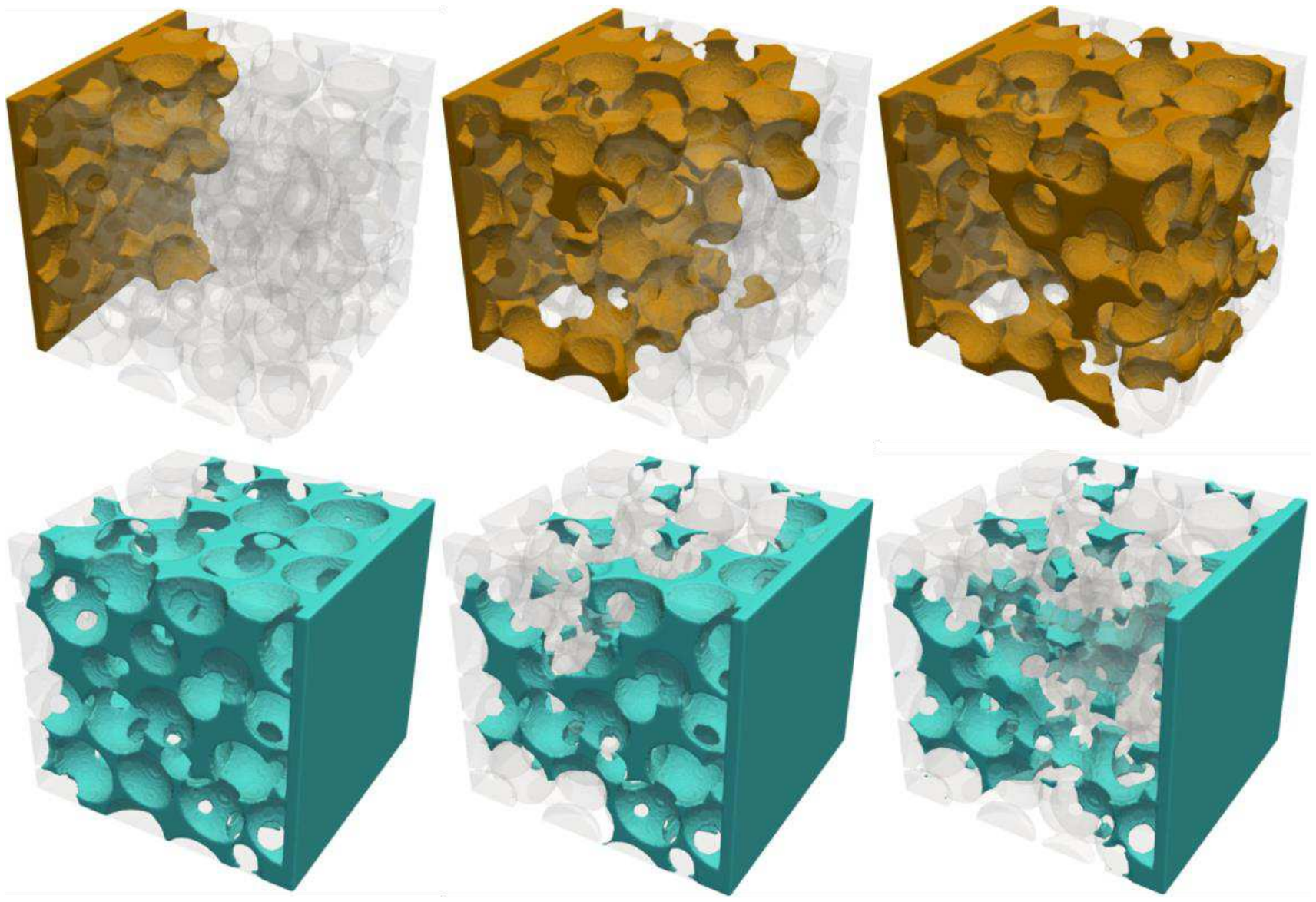

$S_w = 0.874$ $S_w = 0.46$ $S_w = 0.204$

*Figure 6 Snapshots for the primary drainage displacement process in the Finney pack of spheres. The non-wetting fluid (oil) invaded the domain from the left side. The top row of snapshots shows the oil while the bottom row shows the water. Solid was shown in transparent grey in both rows of the snapshots.*

### *4.2. Immiscible fluids displacement in porous rock samples*

We now consider an actual porous rock sample of a Bentheimer sandstone. The dataset representing the geometry was acquired from the Digital Rocks Portal under a project named "*A large scale X-Ray micro-tomography dataset of steady-state multi-phase flow*" [39]. The original size of the dataset was 1950 x 1950 x 10800 with a voxel resolution of 6 µm. Thus, the original dataset size was 11.7mm x 11.7mm x 64.8mm, representing a core sample from a Bentheimer sandstone. A subset with the size of 200 x 200 x 400 (1.2mm x 1.2mm x 2.4mm) was extracted and used in the present simulations. Since this is an actual rock sample representing a heterogeneous porous medium, some image processing was needed. For the sake of numerical efficiency, the isolated pores that do not contribute to the fluid flow were removed from the rock sample. This task was accomplished using Matlab image processing functions such as "*bwareaopen*" [40]. Also, very narrow pores were removed from the sample so that each pore has at least three lattice points in each of the three dimensions. This was achieved using the Matlab function "*imclose*" [41]. A three-dimensional view of the rock sample is shown in *Figure 7*. The porosity of the sample was 20.7 %.

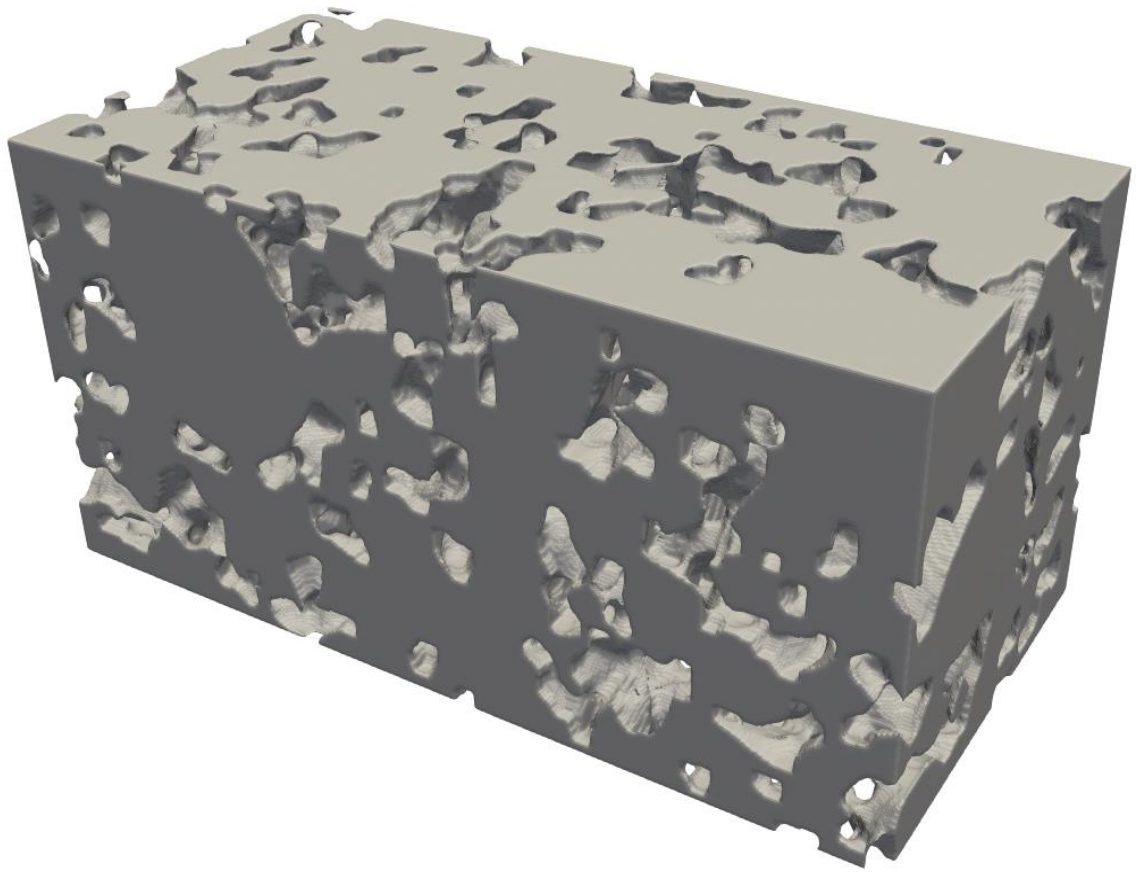

*Figure 7 A three-dimensional view of the Bentheimer sandstone rock sample used in the simulation. The sample size is 200 x 200 x 400.*

We assume that oil is the wetting fluid and water is the non-wetting fluid, which is typical in oil reservoirs [30]. We simulate water displacement by oil in an initially water-saturated rock sample. Such an experiment is widely conducted in SCAL laboratories to generate the $P_c - S_w$ curve of the reservoir rock. We use pressure boundary conditions at both inlet and outlet sides to drive the flow, and solid walls at the boundaries in the y and z directions. This setup matches a typical experiment where the core sample is held in a solid core holder that enforces fluid flow in the x-direction. We use the same technique of allocating pure reservoir regions as described in Sec 4.1. Three cases of different viscosity ratios were simulated, namely M = {5, 25, 100}. The initial density values were set as in Sec. 3.1. The contact angle was set as $\theta \cong 108°$ ( $\xi = 0.2$).

The simulation results are shown in Figure 8. We start with discussing the results from the case of M = 25 in detail and considering it the base case. At zero capillary pressure, there was a considerable amount of oil that invaded the domain even without applying any pressure ($S_o \cong 0.065$). This phenomenon is known as *spontaneous imbibition* [38], and it is physically known to occur for the wetting fluid (oil in this case) due to its preference to wet the solid surface of the rock. Then, as the inlet pressure was gradually increased, the water saturation started decreasing until an irreducible saturation value was achieved despite the increase in the capillary pressure, which in this case was $S_{w,i} \cong 0.29$. Such a residual saturation value is realistic and matches typical values obtained in laboratory experiments [38] [42]. It also assures that the present model has the same advantage of preserving the entrapped fluid volume as described in Sec. 4.1 and in Ref. [16] in realistic porous rock samples and such a complicated porous geometry. We also note that the $P_c - S_w$ curve for this case does not contain any capillary pressure bumps as compared to the results in Ref. [20] because the present geometry did not feature zones with highly varying permeabilities and the viscosity ratio was moderate. Moreover, the shape of the acquired $P_c - S_w$ curve matches the typical curve shapes acquired from such an experiment. Snapshots for the displacement process at different saturations are shown in Figure 9.

At lower viscosity ratio (M = 5), similar features and phenomena such as the spontaneous imbibition and gradual decrease of water saturation against capillary pressure were observed. Also, the residual water saturation value was realistic. However, it could be observed that the less viscous oil was able to invade more regions in the domain at lower capillary pressures. This matches the expected physical behavior of the system since the lower oil viscosity would result in a higher oil permeability. This higher permeability is known to reduce the capillary pressure in the $P_c - S_w$ relationship [43].

At higher viscosity ratio (M = 100), the spontaneous imbibition phenomenon was still observed however the $P_c - S_w$ relationship was not as gradual as the other cases. It could be observed at two water saturations ($S_w \cong 0.67, S_w \cong 0.56$) there are abrupt increases in the capillary pressure values, where an increase in the capillary pressure did not result in a considerable decrease in the water saturation. This phenomenon is known as the capillary pressure bump phenomenon [38]. It was previously reported in the literature that the LBM model could capture this physical phenomenon in 2D [20] and it is reassured in this work that this phenomenon could be captured in 3D as well using the present model. Another feature that could be observed in this case is that the more viscous oil needed higher capillary pressures to invade different regions in the domain. In a reversed manner to what was observed with the less viscous oil, the higher viscosity resulted in lower oil permeability which resulted in an increase in the capillary pressure.

The above observations show that the presented model could be extended to different flow conditions and can capture the physical behaviors of the system and its underlying physics. However, we observed the presence of a checkerboard pattern [6] in the velocity field of some simulations in this geometry for other contact angles (e.g., setting the oil as the non-wetting fluid), which leads to numerical instabilities. A deeper analysis would be required to determine the source of this effect and how to prevent it. The possibility of the formation of the checkerboard pattern in the velocity field could be considered a limitation of the present model in low porosity media.

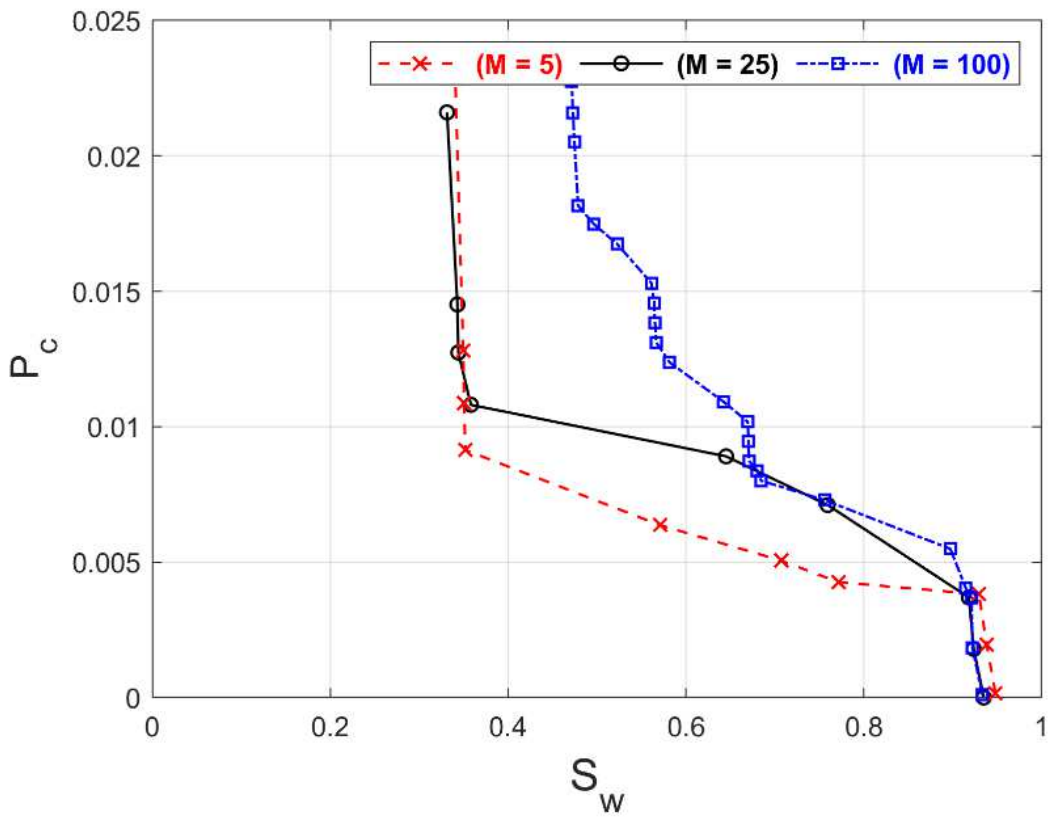

Figure 8 *Capillary pressure* $(P_c)$ *– water saturation* $(S_w)$ *relationship for displacement process in the Bentheimer sandstone rock sample at contact angle* $\theta = 108°$ *and different viscosity ratios.*

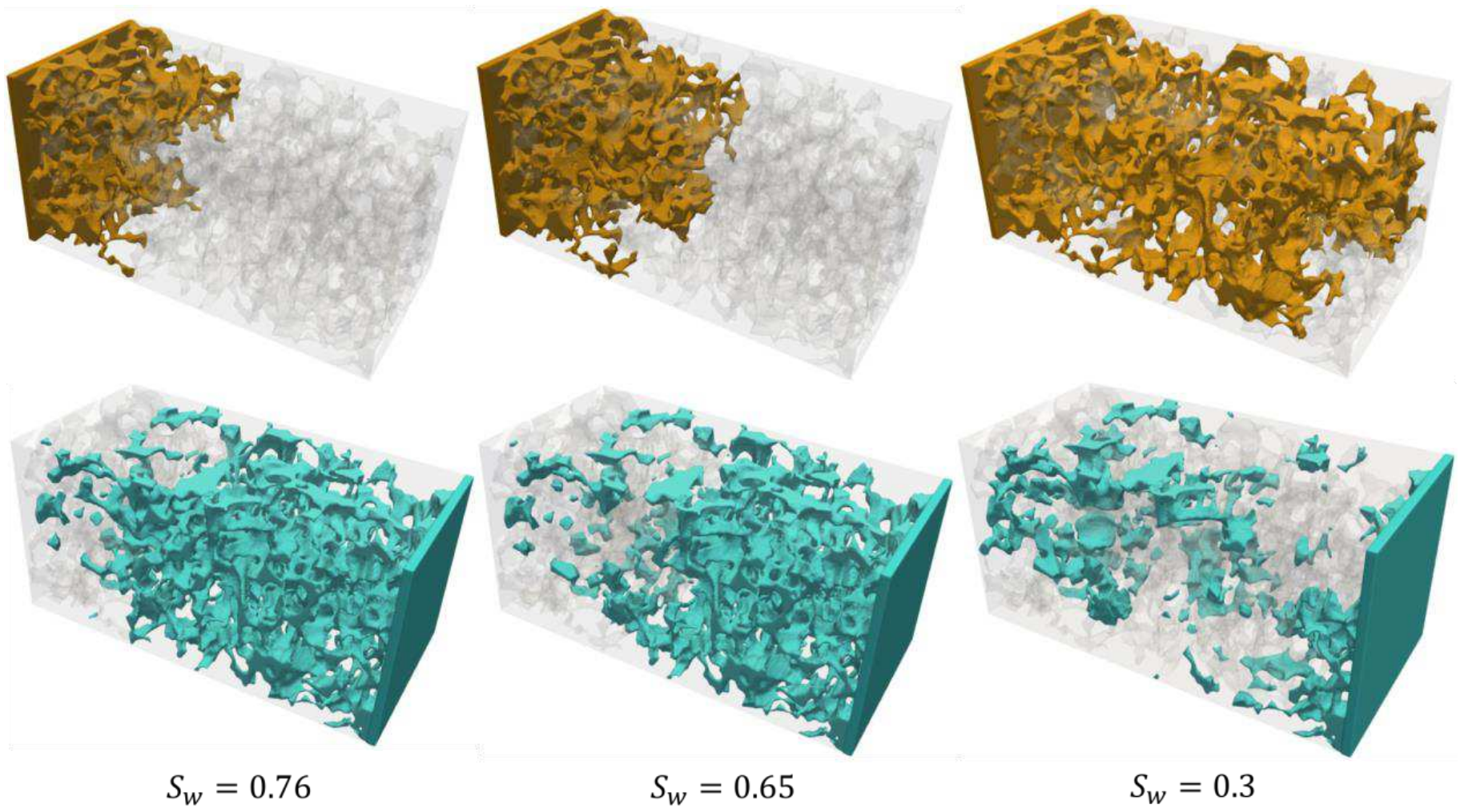

Figure 9 *Snapshots for the displacement process in the Bentheimer sandstone rock sample (M=25). The wetting fluid (oil) invaded the domain from the left side. The top row of snapshots shows the oil, while the bottom row shows the water. Solid was shown in transparent grey in both rows of the snapshots.*

**5. Discussion of the model performance on Graphical Processing Unit (GPU)**

Three-dimensional numerical simulations usually demand significant computational resources due to the amount of data to be processed. They also require massive memory requirements to store all the information needed to carry out the simulation. LBM is naturally a good candidate for massively parallel computing due to its intrinsic parallel nature via many local operations in its algorithm. However, it still needs a significant amount of memory to process relatively large three-dimensional problems. For instance, the simulations carried out in section 4.2 for a realistic sample of a sample of Bentheimer sandstone rock have sample dimensions of 420 x 200 x 200 = 16,800,000 lattice points. At each lattice point, two sets of distribution functions exist that each represent a fluid component. Each set of distribution functions consists of 19 directions representing the used lattice arrangement. To avoid overlapping values and race conditions during the streaming step, we store two copies of each set of the distribution functions. Moreover, at each lattice point, minimum information such as fluid density (both components), velocity vectors in three dimensions, pressure, and lattice point type are stored as well and used during the computations at each time step. Therefore, it should be expected to need a significant amount of memory to store all this information at each lattice point. For single-precision floating-point accuracy, each float number needs four bytes to be saved in the memory. The memory requirements for the different simulation cases are shown below.

We optimized the computational codes to increase its performance, access relevant time and length scales, and acquire results within a reasonable time. We used the Structure of Array (SOA) memory layout starting with the memory layout as it is more suitable for LBM implementation on GPU [44]. In this memory layout, the values of one distribution function direction of all lattice points are placed consecutively in memory. For the multi-component model, we place the values of one distribution function direction for each fluid component consecutively in memory. We implement the streaming step via the pull scheme rather than the push scheme, as it was reported in Ref. [44] to result in better performance.

Since we are concerned with fluid flow through porous media, it was necessary to implement a sparse geometry memory optimization technique to reduce the required information to be stored at each lattice point. This could be theoretically achieved by removing the lattice points that are surrounded by solid points where no data needs to be stored, no equations need to be integrated, and no boundary conditions need to be implemented. For that purpose, we use the technique reported in Ref. [45] for memory tiling, where the computational domain is discretized into equally sized blocks. The blocks that are filled with solid points are removed from memory as they are not needed in the computations. Only the memory blocks containing fluid or fluid-solid interface points are stored. This tiling technique had a significant impact on the computational speed and memory requirement as will be shown.

The simulations presented in this work were carried out using the Nvidia GTX 1660 Ti, which is a general-purpose consumer product GPU installed in a laptop. It should be noted that such a GPU is not dedicated to scientific computing. However, it will be shown that the computational performance was adequate.

The achieved computational speed for the simulations presented in section 4.1 was ~ 450 MLUPS, and the memory requirement for the Finney packing of spheres was 1.85 GigaByte (GB). It should be noted that the memory requirement without the tiling algorithm was 2.82 GB and the computational speed for the same domain was in the range of ~100 MLUPS. This is a reduction in memory requirement by about 35% and an increase in computational speed by about 450%.

The achieved computational speed for the simulations presented in section 4.2 was ~ 650 MLUPS, and the memory requirement for the Bentheimer sandstone rock sample was 2.7 GB. It should be noted that the memory requirement without the tiling algorithm would have been 5.4 GB. Hence, the simulation would require more than the 6 GB of memory available in the used GPU as the operating system needed about 1 GB of memory to operate. This emphasizes the significant impact of the used tiling algorithm on the memory requirement by exploiting the nature of the geometry of the porous medium as the memory requirement was reduced by about 50%. It should also be noted that the porosity of the domain impacted the reduction in the memory requirement. This could be observed as the memory requirement of the Finney packing of spheres was reduced by only 35 % as that domain had a higher porosity.

It should also be mentioned that the present multi-component model requires non-local information in the interaction force calculation step and during the implementation of the wetting boundary condition. This results in non-coalesced memory accesses that reduce the performance. More careful optimizations could be implemented to minimize the non-coalesced memory accesses and increase the performance of the algorithm; however, they would require more effort and careful implementations in the developed codes.

To the authors knowledge, this is the first time the computational performance is reported after the above-mentioned optimizations were combined and implemented for a MC pseudopotential model to reduce the memory consumption and increase the computational performance on GPU. Other published work discussed the computational performance of some variations of the MC model, for instance, Julien Duchateau *et al.* [46] reported the computational speed of the MC model on a single GPU to be 134 MLUPS. M. Januszewski and M. Kostur [47] introduced an open-source fluid simulation package named "Sailfish" and mentioned the computational performance of the MC model where the results varied between 104 MLUPS and 363 MLUPS, depending on the used hardware. However, it should be noted that results from the literature could not be directly compared with the results in this work due to the differences in the model implementation and hardware components.

## 6. SUMMARY

We generalized to 3D simulations our previous improved pseudopotential LBM model and analyzed the fluid displacement in realistic porous media. With this model it is possible to achieve high viscosity ratios, tune the surface tension and suppress the non-physical behaviors of previous models by preserving the volume of trapped fluids. We validated and calibrated the model with several benchmarks and provided an approximated relation to set the contact angle.

We investigated the performance of the model to simulate immiscible fluid displacement in 3D geometries and focused on porous media with realistic geometries, namely a pack of spheres and a Bentheimer sandstone. Fluids with different wettability were simulated to represent other displacement processes such as drainage and imbibition. It could be observed from the results that the present model captures the underlying physics when compared to typical results from laboratory experiments. We also observed that the values of some critical parameters describing the displacement processes were realistic. For example, the irreducible water saturation $S_{w,i}$ was 0.2 for a primary drainage process, while it was 0.29 for an imbibition process.

The numerical performance of the model on GPU was presented as well. We reported the computational speeds and memory requirements of the model for the simulated cases. Also, we mentioned some of the used optimization techniques to boost the computational speed and reduce the memory requirements.

For future work, several points could be addressed to improve the model. For instance, the origin of the checkerboard effect in the velocity field observed in some of the simulations needs to be further investigated and resolved. Also, an outflow boundary condition for the pseudopotential model that works in low porosities is missing in the literature. Such a boundary condition would be very beneficial for mimicking laboratory experiments. Moreover, more useful simulations such as steady-state and unsteady-state relative permeability tests could be carried out. These simulations could mimic the same laboratory tests and enhance the digital SCAL laboratory concept.

## CONFLICTS OF INTEREST

The authors have no conflicts to disclose.

## ACKNOWLEDGMENTS

M. Sedahmed and H. A. Warda thank the High-Performance Computing (HPC) team from Bibliotheca Alexandria (BA) for their support in developing the computer codes used in the present simulations. R. C. V. Coelho and N. A. M. Araújo acknowledge financial support from